\documentclass[%
preprint,
 amsmath,amssymb,
prb,
]{revtex4-2}

\usepackage{graphicx}% Include figure files
\usepackage{dcolumn}% Align table columns on decimal point
\usepackage{bm}% bold math
\usepackage{color}

\usepackage{tocbibind} %for appendix/SI numbering
\usepackage[toc,page]{appendix} %for appendix/SI numbering

\newcommand\figwidtho{3.37in}
\newcommand\figwidtht{6.69in}

\begin{document}

%\preprint{APS/123-QED}

\title{Learned Mappings for Targeted Free Energy Perturbation between Peptide Conformations}% Force line breaks with \\

\author{Soohaeng Yoo Willow}
\affiliation{%
Department of Chemistry, Illinois Institute of Technology, Chicago, Illinois, 60616
}%
\author{Lulu Kang}%
\affiliation{%
Department of Applied Mathematics, Illinois Institute of Technology, Chicago, Illinois, 60616
}%
\author{David D. L. Minh}%
\email{dminh@iit.edu}
\affiliation{%
Department of Chemistry, Department of Biology, and Center for Interdisciplinary Scientific Computation, Illinois Institute of Technology, Chicago, Illinois, 60616
}%


\date{\today}% It is always \today, today,

\begin{abstract}Targeted free energy perturbation uses an invertible mapping to promote configuration space overlap and the convergence of free energy estimates. However, developing suitable mappings can be challenging. Wirnsberger et al. (2020) demonstrated the use of machine learning to train deep neural networks that map between Boltzmann distributions for different thermodynamic states. Here, we adapt their approach to free energy differences of a flexible bonded molecule, deca-alanine, with harmonic biases with different spring centers. When the neural network is trained until ``early stopping'' - when the loss value of the test set increases - we calculate accurate free energy differences between thermodynamic states with spring centers separated by 1~\AA ~and sometimes 2~\AA. For more distant thermodynamic states, the mapping does not produce structures representative of the target state and the method does not reproduce reference calculations.
\end{abstract}
% DM: Removed the lines below because there is too much emphasis on the negative results
% During the training process, the mapping becomes trapped in a local minimum and is unable to reach the dominant configuration space of the targeted thermodynamic state. Thus, our work advances but also pinpoints a weakness of this intriguing free energy method.}

\maketitle

%\tableofcontents

\section{Introduction}

Free energy calculations are a powerful tool that are increasingly used to design materials \cite{TolborgWalsh22} and drugs \cite{ArmacostCournia20}. Accurately calculating free energy differences between a pair of thermodynamic states often entails performing simulations of a multiple intermediate states along a thermodynamic process connecting the states \cite{PohorilleChipot10}. Simulating multiple intermediates promotes overlap between the most important configuration space in neighboring thermodynamic states, a requirement for the convergence of free energy estimates \cite{PearlmanKollman89, WuKofke05, WuKofke05a}.

Circumventing the simulation of these intermediates could improve simulation accuracy and reduce resource requirements. Simulations of intermediates can have problems that degrade the accuracy of free energy estimates. For example, alchemical processes in which particles are created or destroyed are known to suffer from pathologies such as the end-point catastrophe - there can be poor configuration space overlap between states where particles nearly appear or disappear, leading to spurious discontinuities in free energy estimates along a thermodynamic process - and the presence of artificial energy minima \cite{LeeYork20}. Moreover, intermediate states are usually of no particular scientific interest. Avoiding them could lower resource requirements, allowing more scientists to access the tool and reducing the consumption of computing and energy resources.

Targeted free energy perturbation (TFEP) \cite{Jarzynski02} is an approach that can bypass the simulation of intermediate states. Rather than simulating many intermediates, TFEP achieves configuration space overlap via an invertible mapping between the end states. In 2002, \citet{Jarzynski02} formulated TFEP as an extension of the classic free energy perturbation (FEP) identity \cite{Zwanzig54}. In FEP, configurations are sampled from a single state. The free energy is an exponential average of the difference between the potential energy of each sampled configuration in the target and sampled state (Equation \ref{eq:FEP}). On the other hand, instead of using the same configuration for both potential energy evaluations, TFEP uses the potential energy of the \emph{mapped} configuration in the target state. Along with the Jacobian of the mapping, the potential energy of the mapped configuration is incorporated into a generalized work (Equation \ref{eq:generalized_energy_difference}) in the exponential average (Equation \ref{eq:TFEP}).

There were several precedents to the formalism of \citet{Jarzynski02}. In 1985, \citet{Voter85} performed free energy calculations with volume-preserving translations. In 1995, \citet{SeveranceJorgensen95} considered transformations of bond lengths and angles based on harmonic force field parameters as well as the rotation of a dihedral angle. Although they did not explicitly mention Jacobians, all of their example mappings had a Jacobian determinant of unity; the term was not needed to reproduce analytical free energies. TFEP can be considered as a generalization of the celebrated Jarzynski identity \cite{Jarzynski97, Jarzynski97a} that was published in 1997; in the Jarzynski identity, mapping is performed by a nonequilibrium switching process \cite{OberhoferGeissler05}. In 2000, \citet{MillerReinhardt00} proposed that such switching processes can incorporate scaling transformations.

There have been several extensions and generalizations to TFEP. \citet{HahnThen09} implemented the bidirectional estimator that \citet{Jarzynski02} suggested in the original TFEP paper. Soon after \citet{Jarzynski02}, \citet{MengSchilling02} described a multistate generalization of TFEP to arbitrary statistical distributions, which \citet{PaliwalShirts13} applied to molecular systems. In addition to free energies, the multistate generalization is capable of estimating arbitrary expectation values \cite{MengSchilling02, PaliwalShirts13}. Subsequently, TFEP and its extensions have been applied to spin systems and other statistical distributions, but our focus here will be on molecular systems.

Although TFEP is a compelling concept, it has been difficult to apply because of the lack of suitable mapping functions for molecular systems. Mappings have been developed for expanding a cavity in a fluid \cite{Jarzynski02}, inserting a particle into a fluid \cite{HahnThen09}, the interconversion of water models \cite{PaliwalShirts13}, and between crystals at different temperature \cite{TanKofke10} and volume \cite{MoustafaKofke15}. However, the mappings are not broadly generalizable; human intuition and creativity have been required for every distinct application.

While human intelligence is still important, we are now in the age of artificial intelligence. Apropos to the emerging era, a few groups have used deep learning to train maps between molecular systems for TFEP. \citet{WirnsbergerBlundell20}, from the Google subsidiary DeepMind, who also developed AlphaFold \cite{JumperHassabis21}, learned mappings to ensembles of fluids containing a solute with different radii. \citet{DingZhang20} trained mappings between a Boltzmann and tractable distribution - for which the normalized density is known and independent and identically distributed samples can be easily generated - to compute absolute free energies for different conformations of di-alanine and temperatures of deca-alanine. Subsequently, they applied the same approach to a host-guest system to compute binding free energies \cite{DingZhang21a}. \citet{RizziParrinello21} trained neural networks for mapping between two levels of quantum theory for a simple chemical reaction in the gas phase. \citet{WirnsbergerBlundell22} learned mappings between a lattice model with random perturbations to different phases of Lennard-Jones systems.

In recent years, the use of learned mappings in molecular simulation beyond TFEP has been pioneered by Frank No\'{e} and collaborators. In 2019, \citet{NoeWu19} introduced Boltzmann generators, which use a reference state and a learned mapping to generate samples that may be reweighed to the Boltzmann distribution. They demonstrated the method on generating samples of a bistable dimer in fluid (initially described in an paper coauthored by the corresponding author \cite{NilmeierChodera11}) and a small protein. While their publication primarily focused on sampling opposed to free energies, they did report computing free energy differences between independent Boltzmann generators using the average (opposed to the exponential average from TFEP \cite{Jarzynski02}) of the generalized work. Subsequently, \citet{SbailoNoe21} described the use of learned mappings for Monte Carlo moves, as demonstrated in the bistable dimer system. Finally, \citet{InvernizziNoe22} used learned mappings for replica exchange, performing simulations of di-alanine and tetra-alanine. Besides No\'{e} and collaborators, \citet{MahmoudLill22} have developed a hierarchical sampling procedure to generate samples of a medium-sized (106 residue) protein.

Here, we apply learned mappings to free energy differences between different conformations of deca-alanine. Our approach is similar to \citet{WirnsbergerBlundell20} but the system is qualitatively different. \citet{WirnsbergerBlundell20} modeled a cavity in a fluid of neutral monatomic molecules, which has a simpler potential energy function due to the lack of bonded (bond length, bond angle, and torsion) and Coulomb interactions. Our system is similar to  \citet{DingZhang20}, but the approach is qualitatively different. While \citet{DingZhang20} learned mappings between Boltzmann and tractable distributions, we learn mappings between Boltzmann distributions corresponding to different thermodynamic states (different harmonic biases). Mapping to a tractable distribution has an advantage in that the tractable distribution can be selected to have significant configuration space overlap with the unmapped molecular distribution. Moreover, it is generally faster to evaluate the probability density and its derivatives for a tractable distribution than the potential energy of a molecular system. On the other hand, directly mapping between pairs of molecular distributions can permit a subset of coordinates to be perturbed. For processes in which changes are limited to a subset of coordinates, such as those in the focused confinement method \cite{vanderVaartLePhan19, OrndorffvanderVaart20}, such mappings may be particularly fruitful. Hence both approaches to mapping may find separate niches.

The remainder of the paper is as follows. In Section \ref{sec:theory}, we review the theory of TFEP and the loss function. We then describe computational methods (Section \ref{sec:computational_methods}), including the deca-alanine model system, the neural network architecture, training procedure, and free energy estimators. We then report results along with discussion of their implications (Section \ref{sec:results_and_discussion}), and lastly describe our conclusions (Section \ref{sec:conclusions}).

\section{Theory\label{sec:theory}}

\subsection{Targeted free energy perturbation}

Let us first define notation. For a thermodynamic state $A$, a configuration $\bm{x}$ has the equilibrium probability density $\rho_{A}(\bm{x}) = \exp \left[ -\beta U_A(\bm{x}) \right] / Z_A$. In this expression, $\beta$ is the inverse of the temperature and Boltzmann's constant $k_B$, $\beta = \left(k_B T\right)^{-1}$, $U_A(\bm{x})$ is the potential energy of the configuration in state $A$, and $Z_A = \int \exp \left[ -\beta U_A(\bm{x}) \right] d\bm{x}$ is the configurational integral of the state, an integral over all space. Analogous definitions apply to a second thermodynamic state $B$. The objective of our calculation is the Helmholtz free energy difference between states $A$ and $B$, defined as,
\begin{equation}
\Delta F = F_B - F_A = -\beta^{-1} \ln \left( \frac{Z_B}{Z_A} \right).
\end{equation}

In principle, a free energy difference $\Delta F$ may be calculated based on the classic free energy perturbation identity \cite{Zwanzig54},
\begin{eqnarray} \label{eq:FEP}
    \mathbb{E}_A \left[ e^{-\beta \Delta U} \right] = e^{-\beta \Delta F},
\end{eqnarray}
where $\mathbb{E}_A[\cdot]$ represents an expectation value in the state $A$ and $\Delta U(\bm{x}) = U_B(\bm{x}) - U_A(\bm{x})$ is the potential energy difference of the \emph{same} configuration between the two states. (For notational simplicity, the configuration dependence is implicit within the expectation value). Specifically, a free energy difference may be estimated using the sample mean of the observable $\exp \left( -\beta \Delta U(\bm{x}) \right)$ over configurations drawn from state $A$. As exponential averages suffer from finite-sample bias \cite{ZuckermanWoolf04}, better estimation performance may be achieved using samples from both $A$ and $B$ \cite{Bennett76} or from multiple states (including $A$ and $B$) \cite{ShirtsChodera08}. Regardless of the estimator, free energy calculations do not reliably converge to the true values unless there is overlap between the highest-probability regions of configuration space for the pair (or series) of states \cite{PearlmanKollman89, WuKofke05, WuKofke05a}.

\begin{figure}
     \includegraphics[width=\figwidtho]{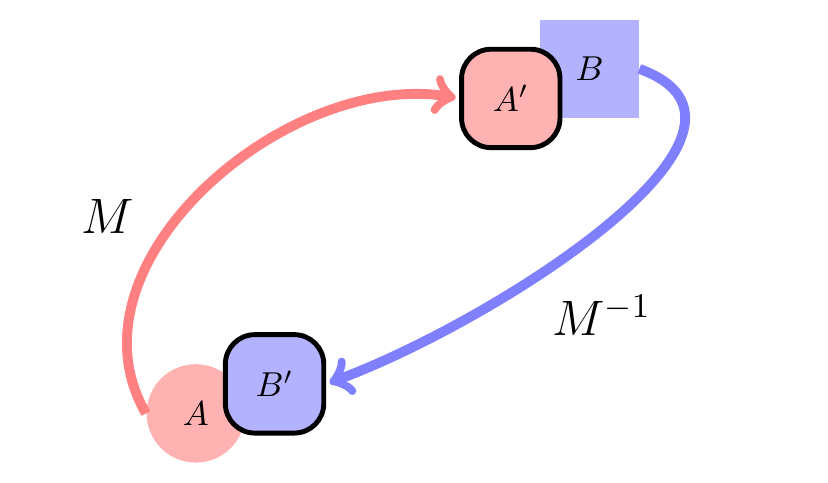}
	\caption{Schematic showing how mapping can improve configuration space overlap. Each shape represents the highest-probability regions of the respective states.
    \label{fig:mapping}}
\end{figure}

 Configuration space overlap may be improved by a bijective (one-to-one) and invertible mapping (Figure \ref{fig:mapping}). A mapping is simply a function that changes a configuration $\bm{x}$ into a new configuration $\bm{x}' = M(\bm{x})$. For an invertible mapping, the inverse function transforms $\bm{x}'$ into $\bm{x}$, $\bm{x} = M^{-1}(\bm{x}')$. Applying a mapping to state $A$ creates a new state $A'$. The probability of the new state is given by the change of variables formula,
\begin{eqnarray}
    \rho_{A^\prime}(M(\bm{x})) & = & \frac{\rho_A (\bm{x})}{|J_{M}(\bm{x})|},
\end{eqnarray}
where $|J_{M}(\bm{x})|$ is the determinant of the Jacobian of the mapping. Similarly, the inverse mapping can be applied to the state $B$ to create a new state $B'$ with density,
\begin{eqnarray}
    \rho_{B^\prime}(M^{-1}(\bm{x})) & = & \frac{\rho (\bm{x})}{|J_{M^{-1}}(\bm{x})|}.
\end{eqnarray}
A mapping can be defined such that $A'$ is close to $B$ and conversely $B'$ is close to $A$ (Figure \ref{fig:mapping}).

Mappings may be incorporated into free energy calculations via a generalization of Equation \ref{eq:FEP}, TFEP \cite{Jarzynski02},
\begin{eqnarray} \label{eq:TFEP}
    \mathbb{E}_A \left[ e^{-\beta \Phi_F} \right] = e^{-\beta \Delta F}.
\end{eqnarray}
Equation \ref{eq:TFEP} replaces the $\Delta U$ in Equation \ref{eq:FEP} with a generalized work,
\begin{eqnarray} \label{eq:generalized_energy_difference}
    \Phi_F (\bm{x}) & = & U_B(M(\bm{x})) - U_A (\bm{x}) - \beta^{-1} \log|J_M(\bm{x})|,
\end{eqnarray}
in which the configurations in state $A$ and $B$ may be different. We refer to this term as work because it is a generalization \cite{OberhoferGeissler05} of the nonequilibrium work from Jarzynski's identity \cite{Jarzynski97, Jarzynski97a}. The work includes a subscript $F$ to denote a ``forward'' mapping from $A$ to $A'$. For the ``reverse'' mapping from $B$ to $B'$, the work includes the subscript $R$,
\begin{eqnarray}
    \Phi_R (\bm{x}) & = & U_A(M^{-1}(\bm{x})) - U_B(\bm{x}) - \beta^{-1} 
    \log | J_{M^{-1}} (\bm{x}) |.
\end{eqnarray}
The designation of a direction of forward or reverse is arbitrary. In the case of an identity mapping, the work is the conventional potential energy difference and Equation \ref{eq:TFEP} reduces to Equation \ref{eq:FEP}. However, a mapping that increases configuration space overlap can improve the convergence of free energy estimates. Indeed, if a mapping perfectly transforms between $A$ and $B$,
\begin{eqnarray} \label{eq:overlap}
    \frac{\rho_{A^\prime}(M(\bm{x}))} {\rho_B(\bm{x})} =  
    \frac{\rho_{B^\prime}(M^{-1}(\bm{x}))}{\rho_A (\bm{x})} = 1,
\end{eqnarray}
then only a single sample is needed to estimate the free energy difference \cite{Jarzynski02, RizziParrinello21}. In addition to the sample mean estimator based on Equation \ref{eq:TFEP}, mappings have also been incorporated into bidirectional \cite{HahnThen09} and multistate \cite{MengSchilling02, PaliwalShirts13} estimators for free energies and expectation values.

\subsection{Loss functions}

While it can be challenging to \emph{design} mappings that improve configuration space overlap and accelerate the convergence of free energy calculations, \citet{WirnsbergerBlundell20} demonstrated that machine learning may be used to \emph{train} mappings. Inspired by configuration space overlap, they developed loss functions for their machine learning models based on a statistical distance between mapped and targeted distributions, the Kullback-Leibler (KL) divergence \cite{Kullback59}. The quality of the forward mapping is quantified by the KL divergence between the mapped density $\rho_{A^\prime}(M(\bm{x}))$ and the target density $\rho_B (\bm{x})$,
\begin{eqnarray}
    D_\mathrm{KL}[\rho_{A^\prime} || \rho_B] 
    & = & \beta (\mathbb{E}_{A} [\Phi_F] - \Delta F) \label{eq:KL}.
\end{eqnarray}
Conversely, the quality of the reverse mapping is described by the KL divergence between $\rho_{B^\prime}(M^{-1}(\bm{x}))$ and $\rho_A (\bm{x})$,
\begin{equation}
    D_\mathrm{KL} [\rho_{B^\prime} || \rho_A] = \beta (\mathbb{E}_B [\Phi_R] + \Delta F).
\end{equation}
As $\beta$ and $\Delta F$ are constants with respect to mapping, they do not need to be included in loss functions. Hence, a loss function suitable for training only the forward mapping is,
\begin{equation}
\mathcal{L}_F = \mathbb{\bar{E}}_A[\Phi_F],
\end{equation}
where $\mathbb{\bar{E}}_A[\cdot]$ is a sample mean estimate of an expectation value in state $A$. 
\citet{WirnsbergerBlundell20} also introduced and recommended a bidirectional loss function,
\begin{equation} \label{eq:loss}
\mathcal{L} = \mathbb{\bar{E}}_A[\Phi_F] + \mathbb{\bar{E}}_B[\Phi_R],
\end{equation}
which we use in the present work.

\section{Computational Methods\label{sec:computational_methods}}

\subsection{Model system}

Alanine deca-peptide (ACE-(ALA)$_9$-NME) was selected as a model system. The system was modeled in gas phase at room temperature, where it is a stable $\alpha$ helix. The initial structure was modified from \citet{NguyenMinh16} by hand to include the end caps ACE and NME. The system has 102 atoms. Parameters from the AMBER ff14SB force field \cite{MaierSimmerling15} were used. Harmonic restraints were added to the carbon of the first peptide bond and the nitrogen atom of the last peptide bond, leading to the total potential energy, \begin{equation}
    U_\lambda(x) = U_0 (x) + k || \bm{x}_C - \bm{x}_C^0 ||^2 + k || \bm{x}_N - \bm{x}_N^\lambda ||^2,
\end{equation}
where $U_0 (x)$ is the standard gas-phase AMBER potential energy. 
In the harmonic potential, $k = 50$ kJ/mol/\AA$^2$ is the force constant, 
$\bm{x}_C$ and $\bm{x}_N$ are coordinates of the restrained carbon and nitrogen atoms, respectively, 
and $|| \cdot ||$ represents the Euclidean norm, the length of the vector.
The spring center $\bm{x}_C^0$ is at the origin and $\bm{x}_N^\lambda = (0, 0, \lambda)$, where $\lambda \in \left\{ 14, 14.2, 14.4, ..., 20.6, 20.8, 21 \right\}$ \AA. 

For each $\lambda$, an ensemble of structures was generated with a molecular dynamics simulation. Isothermal molecular dynamics simulations were performed using the Langevin integrator (LangevinIntegrator) in OpenMM 7.7 \cite{EastmanPande17} at a temperature of $T=300$ K with a time step of 1 fs for a duration of 20 ns. The last 10000 configurations (every 1 ps for the last 10 ns) were saved.

\subsection{Neural network architecture}

Mappings were performed using real-valued non-volume preserving (real NVP) transformations \cite{DinhBengio17} between Cartesian coordinates. Real NVP uses a deep neural network composed of a stack of simple bijections that include scaling and translation. It is especially well-suited for TFEP because it is stably invertible and because the Jacobian of the transformation can be efficiently computed.

\begin{figure}
\begin{tabular}{c c}
    \large{(a)} \\ & \includegraphics[width=5in]{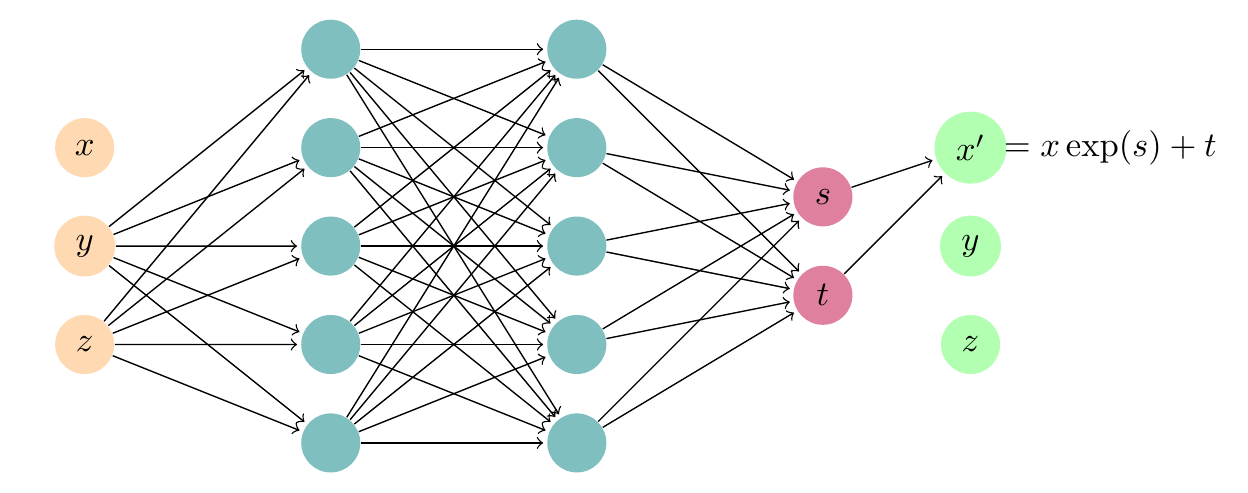} \\
   \large {(b)} \\ & \includegraphics[width=5in]{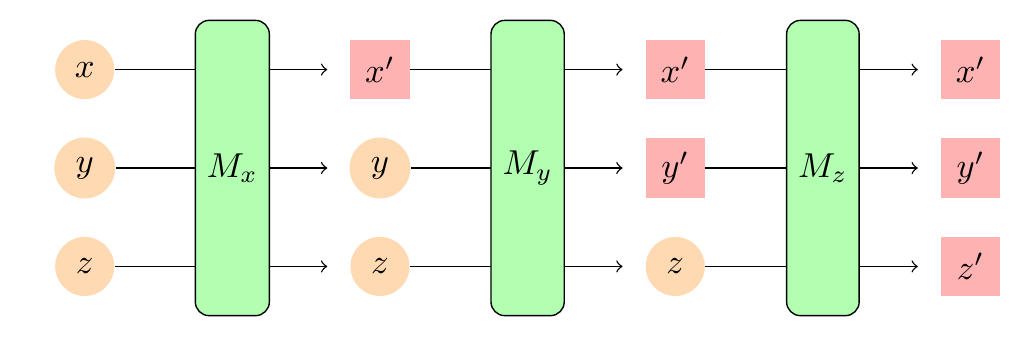} 
\end{tabular}
	\caption{Schematic of the neural network architecture. (a) In the first coupling layer $M_x$,  $y$ and $z$ are left unchanged and the learnable parameters $\theta$ for $s$ and $t$ functions are trained to transform the $x$ coordinates. (b) The entire real NVP network consists of $M = M_z \circ M_y \circ M_x$.
    \label{fig:nn}}
\end{figure}

To adapt real NVP to molecular systems in Cartesian coordinates, we used separate coupling layers for $x$, $y$, and $z$ dimensions (Figure \ref{fig:nn}). The first coupling layer $M_x$ transforms the $x$ coordinates while leaving the $y$ and $z$ coordinates invariant, 
\begin{eqnarray}
M_x (\{x,y,z\}) = \left\{ x^\prime = x e^{s(y,z)} + t(y,z), y, z \right\}
\end{eqnarray}
where $s$ and $t$ are functions encoded in the neural network. The inverse of this mapping is 
$M_x^{-1} (\{x^\prime, y, z\}) = \left\{ x = \left[x^\prime - t(y, z) \right]  e^{-s(y,z)}, y, z \right\}$. 
Jacobian determinants of the mappings $M_x$ and $M_x^{-1}$ are,
\begin{eqnarray}
    \log|J_{M_x}|  =  \sum_{i=1}^{N_\mathrm{atoms}} s_i(y,z) \\
    \log|J_{M_x^{-1}}|  =  -\sum_{i=1}^{N_\mathrm{atoms}} s_i(y,z),
\end{eqnarray}
where the sum is over the atoms and the limit the total number of atoms $N_\mathrm{atoms}$. Similarly, the second coupling layer $M_y$ and the third coupling layer $M_z$ transform the $y$ and $z$ coordinates with $(x, z)$ and $(x,y)$ invariant, respectively. Each coupling layer $M_k: \mathbb{T}^{3N} \rightarrow \mathbb{T}^{3N}$ used the same architecture but different learnable parameters $\theta_k$. The coupling layers consisted of 10 hidden layers with 128 dimensions (compared to 102 atoms).

The overall mapping $M$ of the real NVP transformation is given as
\begin{eqnarray}
    M  =  M_z \circ M_y \circ M_x,
\end{eqnarray}
where $\circ$ means that the output of the right side is input for the left side. Its inverse is,
\begin{eqnarray}
    M^{-1}  =  M_x^{-1} \circ M_y^{-1} \circ M_z^{-1}.
\end{eqnarray}
The Jacobian determinant of the mapping $M$ and the inverse mapping $M^{-1}$ are,
\begin{equation}
    \log |J_M| =  \sum_{\nu \in \{x,y,z\}} \log |J_{M_\nu}|, \\
    \log |J_{M^{-1}}| =  \sum_{\nu \in \{x,y,z\}} \log |J_{M_\nu^{-1}}| 
\end{equation}

The mapping was implemented in JAX \cite{jax2018github}. 

\subsection{Training}

The real NVP map was initialized as an identity map. This was achieved by initializing the transformation $t$ and scaling $s$ factors at zero, such that $x' = x \exp (0) + 0 = x$. Randomness in initial conditions was introduced by drawing all other initial parameters from the standard normal distribution. 
% DM: I don't quite get how to do this, since s and t are outputs from the NN.
% SW: From this simple matrix multiplication: O (output\_dim, h\_dim) @ H$_2$ (h\_dim, h\_dim) @ H$_1$ (h\_dim, input\_dim) @I (input\_dim) = ST(output\_dim), we gave the random parameters to H$_k$, but assigned the zero matrix to O as the initial matrix. Eventually, we have the zero vectors S and T.

The real NVP neural network was trained by minimizing the value of the loss function (Equation \ref{eq:loss}). As the loss function (Equation \ref{eq:loss}) is based on the molecular mechanics energies of mapped configurations, we implemented the AMBER force field in JAX. This implementation is freely available at \url{https://github.com/swillow/jax_amber}. To ensure that mapped $\bm{x}_N$ are similar to $\bm{x}_N^\lambda$, we used $k=200$ kJ/mol/\AA$^2$ instead of $k=50$ kJ/mol/\AA$^2$ in the loss function. 
Data were divided into a training set (80\%) and a test set (20\%). 
Out of the last 10 ns of simulation, 2000 configurations (from 10 ns to 12 ns) were designated as test sets while the remaining 8000 configurations (from 12 ns to 20 ns) were designated as training sets. The training loss was minimized using the adaptive moment estimation (Adam) optimizer \cite{KingmaBa17} with a learning rate of 1.0e-4. 

% DM: For future reference, this way of splitting the training and test set may make them more different from each other and make overfitting a bigger problem.

We performed different amounts of training in initial and later calculations. Initially, training was performed for 150000 steps. We compared results from mappings obtained after different amounts of training:
\begin{enumerate}
\item Early stopping, with the minimum loss of the test set.
\item L $\sim$ 0, where the loss of the training set was closest to zero.
\item Complete, after all 150000 training steps.
\end{enumerate}
Subsequently, after observing superior free energy estimation from early stopping, we stopped training once the loss of the test set in the last 200 steps was over $3.0~k_BT$ greater than the minimum.

\subsection{Free energy estimation}

Reference free energies as a function of $\lambda$ were calculated the multistate Bennett acceptance ratio (MBAR) estimator \cite{ShirtsChodera08} without mapping. They made us of the entire ensemble of equilibrated structures and energies. Mapped free energies were estimated based on data from pairs of $\lambda$ using the generalization of the Bennett Acceptance Ratio \cite{Bennett76} described by \citet{HahnThen09}. Thus, the mapped calculations were based on a small subset of the structures and energies used in the reference calculations.

\section{Results and Discussion\label{sec:results_and_discussion}}

\subsection{The free energy landscape of deca-alanine has multiple barriers}

\begin{figure}
	\includegraphics[width=\figwidtho]{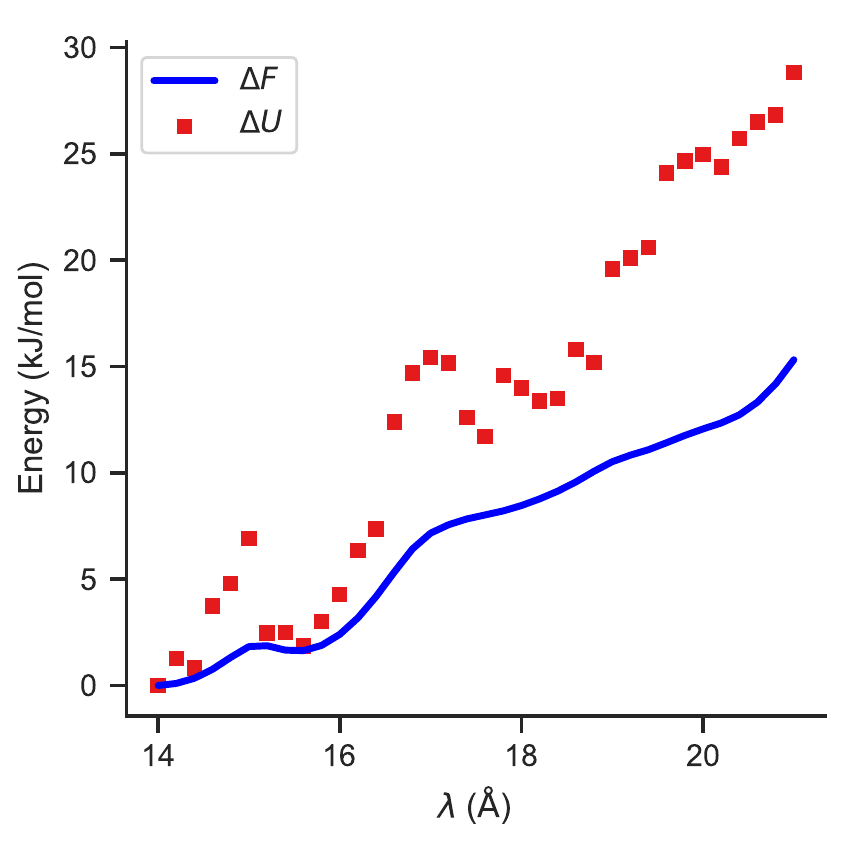}
	\caption{The reference free energy surface and average potential energy difference compared to the state with $\lambda = 14$ \AA.
    \label{fig:FE}}
\end{figure}

The free energy of deca-alanine as as function of the spring position generally increases between $\lambda = 14$ and $\lambda = 21$ \AA~ (Figure \ref{fig:FE}). There are local minima near 14, 16, and 18 \AA~ that are separated by barriers. The barriers correspond to local peaks in the potential energy, suggesting that crossing them corresponds to breaking intramolecular interactions such as hydrogen bonds. As the potential energy barriers are larger than the free energy barriers, they are partially compensated for by entropic increases.

\subsection{Overlap decreases as $\lambda$ separation increases}

While there may be some overlap between thermodynamic states with $\lambda$ separated by 1 \AA, there does not appear to be any overlap between states where it is separated by 2 \AA~ or more ($\Delta \lambda \geq 2$ \AA). The overlap between thermodynamic states may be evaluated based on the histograms of $\Phi_F$ and $-\Phi_R$. Without mapping (or an identity map), the work is simply the potential energy difference of the same configuration in the pair of states. For both separations, distributions of the potential energy difference are unimodal. For $\Delta \lambda = 1$ \AA, there is some overlap between the tails of the distributions of $\Phi_F$ and $-\Phi_R$. For $\Delta \lambda = 2$ \AA, the distributions of potential energy differences are broader and further separated such that there is no evident overlap. Due to the lack of overlap between these distributions, the free energy difference cannot be accurately calculated (without a mapping that improves overlap).

\begin{figure}
     \includegraphics[width=\figwidtho]{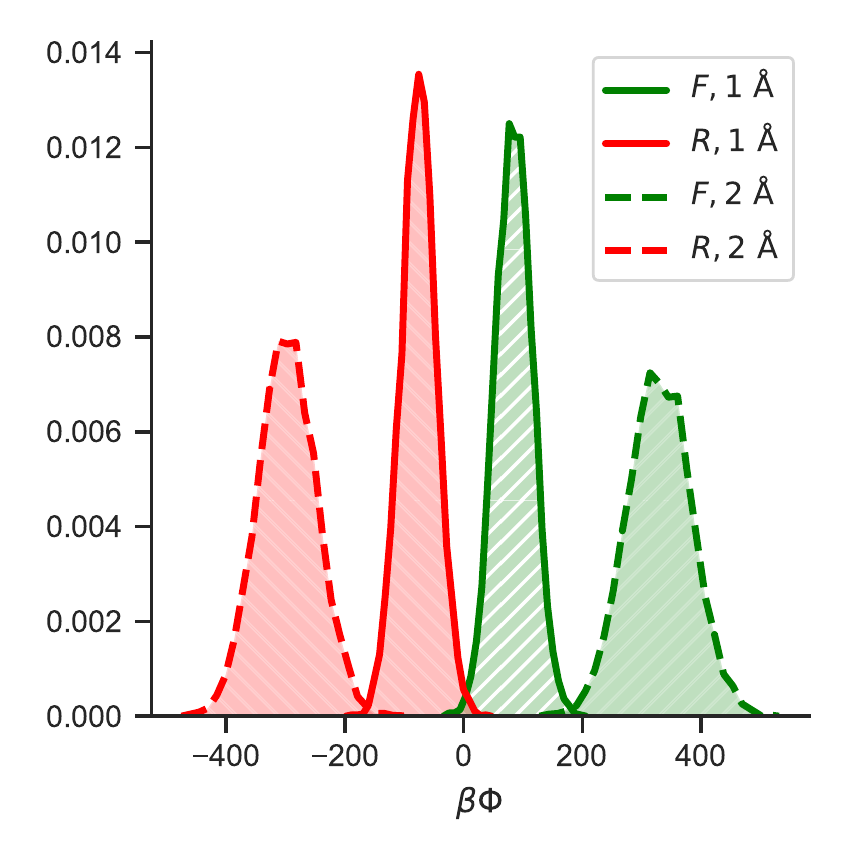}
	\caption{
 Normalized histograms of work without mapping for $\lambda_A = 20$ \AA~and $\lambda_B = 21$ \AA~(solid line) and for $\lambda_A = 19$ \AA~and $\lambda_B = 21$ \AA~(dashed line). For both pairs, $\beta \Phi_F$ are colored green with forward diagonal hatches and $-\beta \Phi_R$ are colored red with backwards diagonal hatches.
    \label{fig:phi}}
\end{figure}

Even for $\Delta \lambda \le 2$ \AA, our unmapped work distributions are more distinct from one another than those reported in \citet{WirnsbergerBlundell20}. \citet{WirnsbergerBlundell20} focused on a solute in a Lennard-Jones fluid, in which the states $A$ and $B$ were generated by changing the solute radii from $R_A = 2.5974 ~\sigma$ to $R_B = 2.8444 ~\sigma$. For $\sigma=3.15$ \AA, the radius of the solute in state $A$ is 8.18 \AA ~and that of the solute in state $B$ is 8.96 \AA, for a difference of 0.78 \AA. The distance between unmapped work distributions was about $\beta (\mathbb{E}_A[\Phi_F] + \mathbb{E}_B[\Phi_R]) \sim 40$. In contrast, we observe $\beta (\mathbb{E}_A[\Phi_F] + \mathbb{E}_B[\Phi_R]) \sim 200$ and 600 for 1 \AA~ and 2 \AA, respectively.

\subsection{Loss values for the training and test set diverge after overfitting}

\begin{figure}
     \includegraphics[width=\figwidtho]{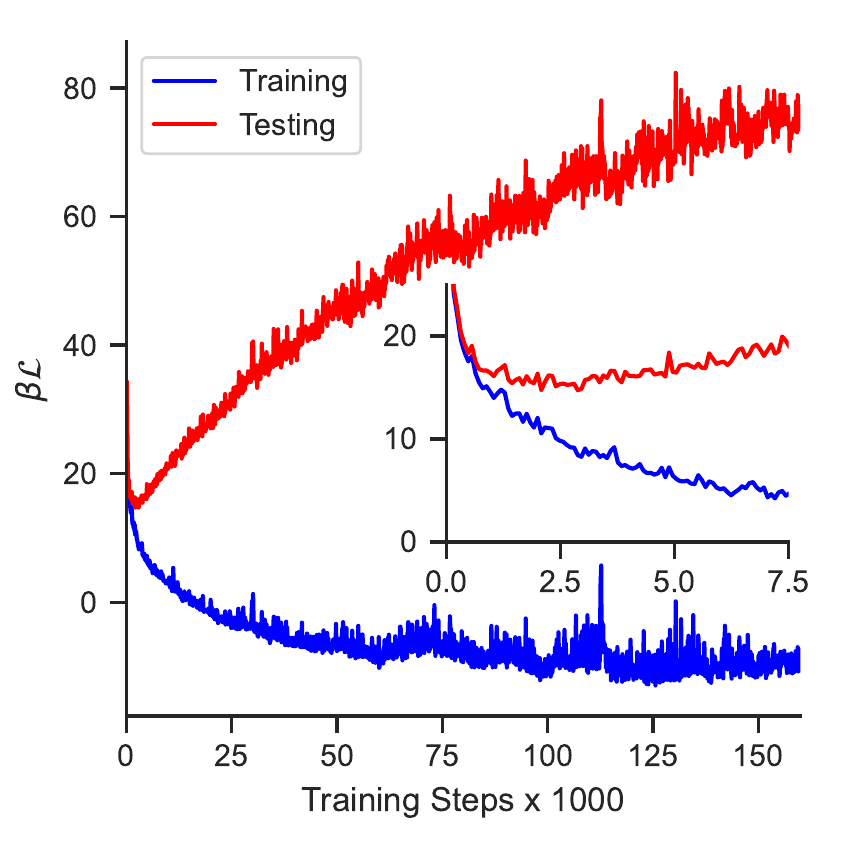}
	\caption{Loss values for the mapping between states with $\lambda_A = 20$ \AA ~and $\lambda_B = 21$ \AA ~ for training (blue) and test (red) data. The inset is a subset of the range emphasizing the minimum of the test loss.
    \label{fig:loss}}
\end{figure}

In all of the learned mappings, the loss values for the training and test set exhibit different behavior from other another (Figure \ref{fig:loss}). At the start of training, both loss values quickly decrease. With additional training steps, the training loss continues to trend downward and even becomes negative. On the other hand, the test loss approaches a minimum and sharply increases after around 2000 steps. 

A deviation between the training and test loss is common in machine learning, including the learned mapping in \citet{WirnsbergerBlundell20} and \citet{RizziParrinello21} - and is an indicator of overfitting to the training set. For this reason, \citet{WirnsbergerBlundell20} adopted an early stopping criterion. 
% SW: If possible, please remove the discussion based on \citet{RizziParrinello21}. They just mapped between PM6 and MP2. In short, both states should have the same (similar) potential energy surface. Except their fancy words and sentences, I don't think that discussing how much hard to mapping between similar PES's with focusing only one bond distance is scientific observation.}
% DM: They also claim to prove the importance of a separate training and evaluation set. I am asking Lulu to think more about this.
\citet{RizziParrinello21} suggested that overtraining leads to a systematic bias in the free energy estimate and thus they estimated free energies based on evaluation sets separate from their training sets.

\subsection{Mapping increases the overlap between forward and reverse work distributions}

\begin{figure}
     \includegraphics[width=\figwidtho]{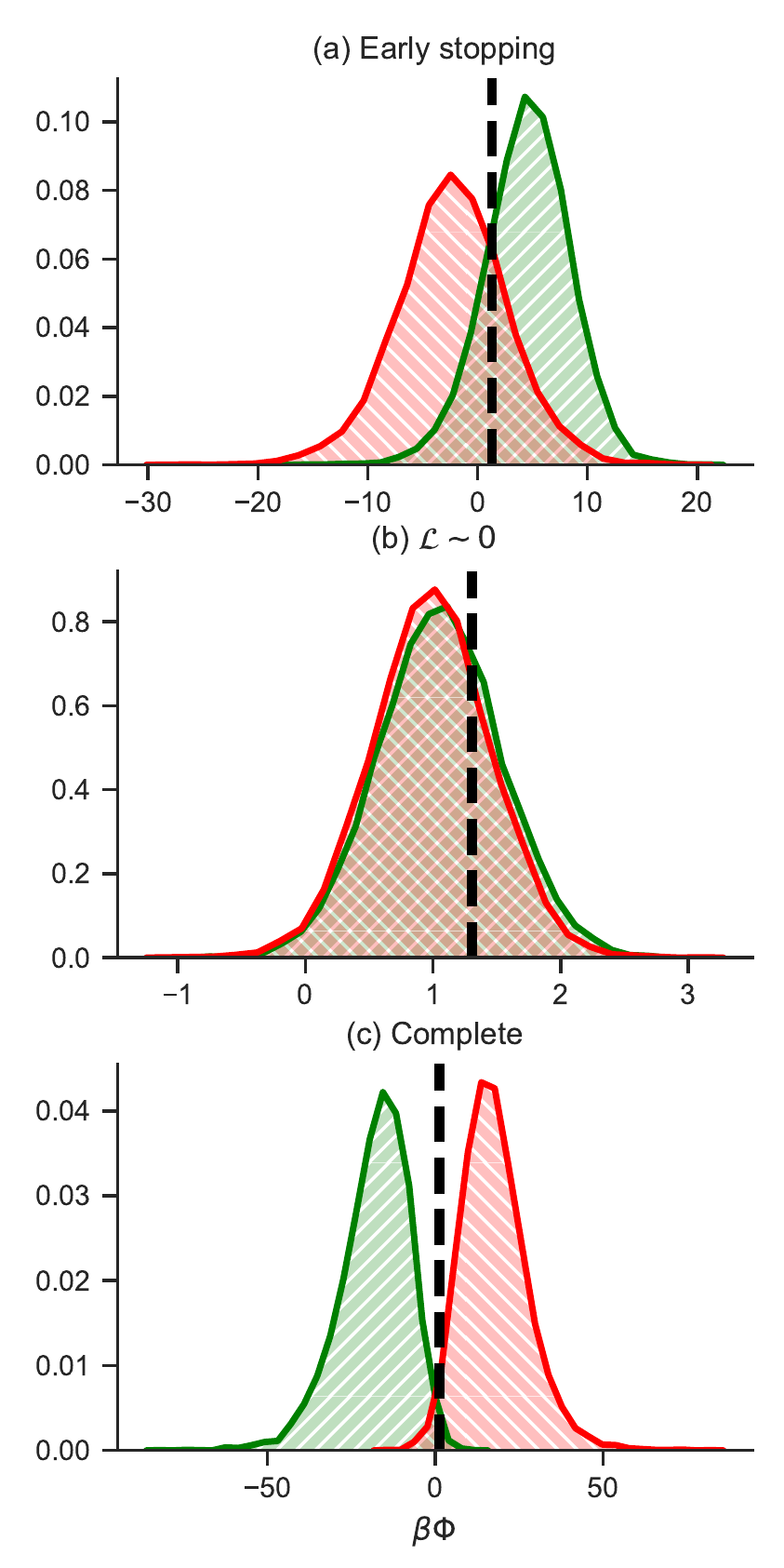} 
\caption{\label{fig:mapped_phi}Effect of stopping criterion on work distributions for training data between $\lambda_A = 20$ \AA~and $\lambda_B= 21$ \AA. Normalized histograms are shown for mapping functions from (a) early stopping, (b) when the loss is closest to zero, and (c) after 150000 training steps. $\beta \Phi_F$ are colored green with forward diagonal hatches and $-\beta \Phi_R$ are colored red with backwards diagonal hatches. The reference free energy difference $\Delta F$ obtained via the MBAR estimator is represented by the vertical dashed line. A similar figure for the testing data is available as Figure S1 in the Supplementary Material.}
\end{figure}

With early stopping, mapping improves overlap between the work distributions and the feasibility of free energy estimation (Figure \ref{fig:mapped_phi}a). At this mapping, which provides a minimum loss for the test set, there is considerable overlap between the distributions of $\Phi_F$ and $-\Phi_R$. Trends in the average work and free energy difference are consistent with applying Jensen's inequality to the expectation values in the TFEP expression, Equation \ref{eq:TFEP}. We anticipate that $\mathbb{E}_A[\Phi_F] > -\beta^{-1} \ln \mathbb{E}_A[e^{-\beta \Phi_F}] = \Delta F$. Similarly, $\mathbb{E}_B[\Phi_R] > -\Delta F$. Hence, expectation values of the work bound the free energy according to $-\mathbb{E}_B[\Phi_R] < \Delta F < \mathbb{E}_A[\Phi_F]$. While these inequalities are true for expectation values, they do not necessarily hold for estimates based on finite sample sizes, which are not only imprecise but systematically biased \cite{ZuckermanWoolf04}. With early stopping, however, the order of expectation values and the free energy follows the anticipated trend. Consequently, the free energy difference, which is at the value of $\Phi$ in which the densities are equal \cite{HahnThen09}, is straightforward to identify. However, further training leads to unexpected trends in the distributions of generalized energies. 

When the loss is close to zero, we observe that both estimated averages are less than the free energy difference, $\mathbb{\hat{E}}_B [-\Phi_R] \sim \mathbb{\hat{E}}_A[\Phi_F] < \Delta F$ (Figure \ref{fig:mapped_phi}b). With a perfect mapping, we would anticipate that $\mathbb{E}_B [-\Phi_R] = \mathbb{E}_A[\Phi_F] = \Delta F$. The deviation of the expectation values from the free energy difference demonstrates that perfect mapping is not achieved. The potential energy distributions of the mapped configurations mostly overlap with the targeted (unmapped) potential energy distribution (Figure \ref{fig:mapped_properties} of the Supplementary Material). This overlap between potential energy distributions shows that the source of the observed inequality is from the Jacobian; $\log |J_M| \neq 0$ and $\log |J_{M^{-1}}| \neq 0$.

Finally, at the end of training, when the training loss values are negative ($\mathcal{L} < 0$), the forward work values are actually much smaller than the reverse work values. As shown in Figure \ref{fig:mapped_phi} (c), the order of averages and free energies is flipped such that $\mathbb{\hat{E}}_A[\Phi_F] \ll \Delta F \ll \mathbb{\hat{E}}_B [-\Phi_R]$. This flipping occurs because mapped energies $U_B (M(\bm{x}_A))$ and $U_A(M^{-1}(\bm{x}_B))$ are lower than the targeted energies $U_B(\bm{x}_B)$ and $U_A(\bm{x}_A)$ (Figure \ref{fig:mapped_properties} of the Supplementary Material). Molecular mechanics potential energies, especially Lennard-Jones repulsion terms, are very sensitive to small changes in coordinates; there are many opportunities for energy optimization relative to samples from the Boltzmann distribution. While these low energies demonstrate that the optimizer is working well, the generated structures are not representative of the target distribution. Moreover, the overlap between work distributions is reduced compared to when the loss is close to zero.

% Because of problems that arise with extended training, the selected bidirectional loss function may not be the ideal choice for TFEP, or even for molecular structures.
% 
% DM: How is this?
% SW: I think that it is not a problem of bidirectional problem. 
% It is a good example why we should use 'early stopping' to avoid overfitting (overtraining) problems.  

Except with early stopping, trends in the work distributions of the test set do not match that of the training set (Figure \ref{fig:mapped_phi_test} in the Supplementary Material.) With early stopping, the training and test set have comparable work distributions. When $\mathcal{L} \sim 0$, the work distributions for the test set are similar to the distributions after early stopping; $\rho(\beta \Phi_F)$ and $\rho(-\beta \Phi_R)$ of the test set do not have nearly complete overlap. For complete training, the anticipated order of averages and free energies $ \mathbb{\hat{E}}_B [-\Phi_R] < \Delta F < \mathbb{\hat{E}}_A[\Phi_F]$ is preserved, but the distributions are very broad. The corresponding large test loss indicates that there is overfitting to the training set.

\subsection{Mapping enables accurate free estimation for smaller $\Delta \lambda$}

\begin{figure}
\includegraphics[width=\figwidtho]{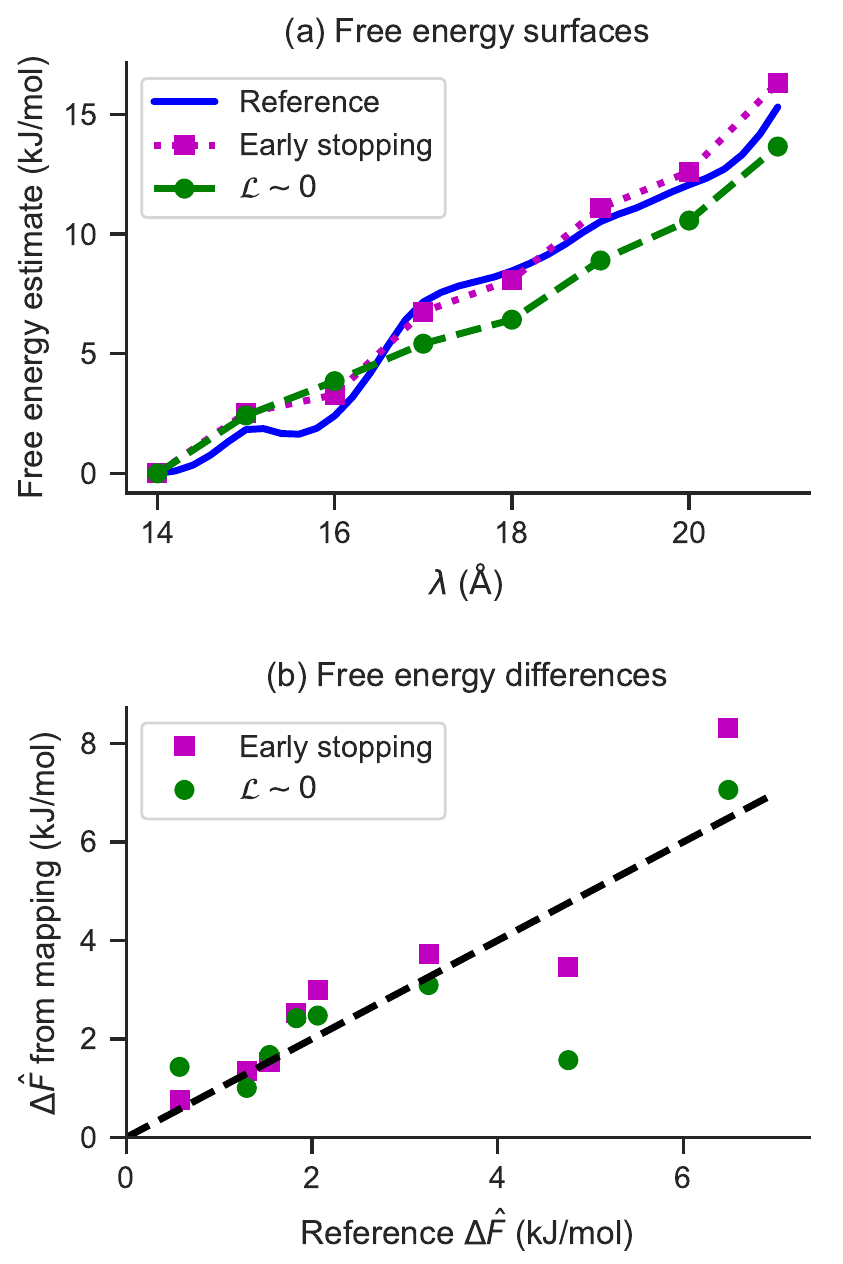}
\caption{Comparison of free energy estimates for a separation of 1~\AA. Mapped free energy differences were computed between $\lambda_A$ and $\lambda_B = \lambda_A + 1$~\AA~for $\lambda_A \in \{ 14, 15, ..., 20 \}$~\AA. (a) Free energy surfaces were computed as a cumulative sum. (b) Individual mapped $\Delta \hat{F}$ were compared to reference values. In both panels, estimates were computed with maps from early stopping (magenta squares) and $\mathcal{L} \sim 0$ (green circles).
\label{fig:dF}}
\end{figure}

For most pairs of states where $\Delta \lambda = 1$ \AA, TFEP is able to accurately reproduce reference free energy differences. Mapped estimates are most accurate between pairs of states with $\Delta F$ less than 2 kJ/mol, less accurate when $\Delta F$ is between 2 and 4 kJ/mol, and least accurate for $\Delta F$ greater than 4 kJ/mol (Figure \ref{fig:dF}b). 

Among stopping criterion, TFEP is more reliable when using a learned mapping with early stopping. When the loss value is near zero, TFEP performance is sometimes improved over early stopping, but in several cases with large free energy differences, the estimate significantly deviates from the reference value. When $-\mathbb{E}_B[\Phi_R] = \mathbb{E}_A[\Phi_F] = \Delta F$, we would expect that the free energies are easy to predict. However, we have already shown (Figure \ref{fig:mapped_phi}b), that the learned mappings do not provide perfect mappings and that there is overfitting to the training set.

Differences between stopping criterion are more evident for free energy surface reconstruction (Figure \ref{fig:dF}a). With early stopping, the estimated free energy surface reproduces the local minimum near 16~\AA~ and the barrier between that state and at 18~\AA. Finer details in the free energy surface are not evident because the mapped estimate was performed at intervals of 1 \AA~opposed to 0.2 \AA. In contrast to the accurate surface reconstruction using the mapping from early stopping, when the training loss is near zero, the estimated free energy surfaces has few features.

These results demonstrate the ability to achieve configuration space overlap and accurate free energy estimation through learned mapping, opposed to through sampling of additional intermediate states. Comparable free energy differences are obtained using a fifth of the number of samples between adjacent $\lambda$ values. While the mapped estimation procedure incurs additional costs in training and is more complex, it may lead to an overall reduction in computational cost.

Using early stopping, reasonable mapped free energy estimates were also obtained for select pairs of states for which $\Delta \lambda = 2$. For $\lambda_A = 18$~\AA~and $\lambda_B = 20$~\AA, the mapped estimate based on early stopping (3.1 kJ/mol) is better than the $\mathcal{L} \sim 0$ estimate (7.4 kJ/mol) at reproducing the reference free energy (3.6 kJ/mol). For $\lambda_A = 19$~\AA~and $\lambda_B = 21$~\AA, the early stopping estimate (5.5 kJ/mol) is also closer to the reference free energy (4.8 kJ/mol) than the $\mathcal{L} \sim 0$ estimate (6.6 kJ/mol). These pairs of states are not separated by potential energy barriers (Figure \ref{fig:FE}).

\subsection{There is poor mapping for more distant states}

In contrast to the accurate performance for smaller $\Delta \lambda$, TFEP does not provide accurate free energy differences for pairs of states in which the spring center is separated by 4 \AA~(Table \ref{tab:dF4}). Not only are TFEP results significantly different from the reference calculation, the two independent training replicates diverge from one another. This suggests that the optimization leads to maps that are quite different from one another. Thus, these maps are at local opposed to global minima of the loss function. To pinpoint the causes and effects of poor mapping we further analyzed the work, potential energy, and structural distributions of the mapping between $\lambda_A = 14$ \AA ~and $\lambda_B = 18$ \AA~.

\begin{table}
\caption{\label{tab:dF4}
Comparison of TFEP and MBAR free energy estimates for 4~\AA~separations. TFEP estimates were based on learned mappings with two independent trainings.}
\begin{tabular}{|c|c | c | c|}
\hline
  $(\lambda_A,\lambda_B)$ & $\Delta F_\mathrm{ref} $  &  $\Delta F_\mathrm{TFEP}$ (Training 1) &  $\Delta F_\mathrm{TFEP}$ (Training 2) \\
  \hline
   (14, 18)  & 8.461 & -20.419 & -35.181 \\
   (15, 19) & 8.691 & -25.454 & -11.418 \\
   (16, 20) & 9.269 & 11.294 & 7.835 \\
   (17, 21) & 8.154 & 4.757 & 17.162 \\
   \hline
\end{tabular}
\end{table}

\begin{figure}
\includegraphics[width=\figwidtho]{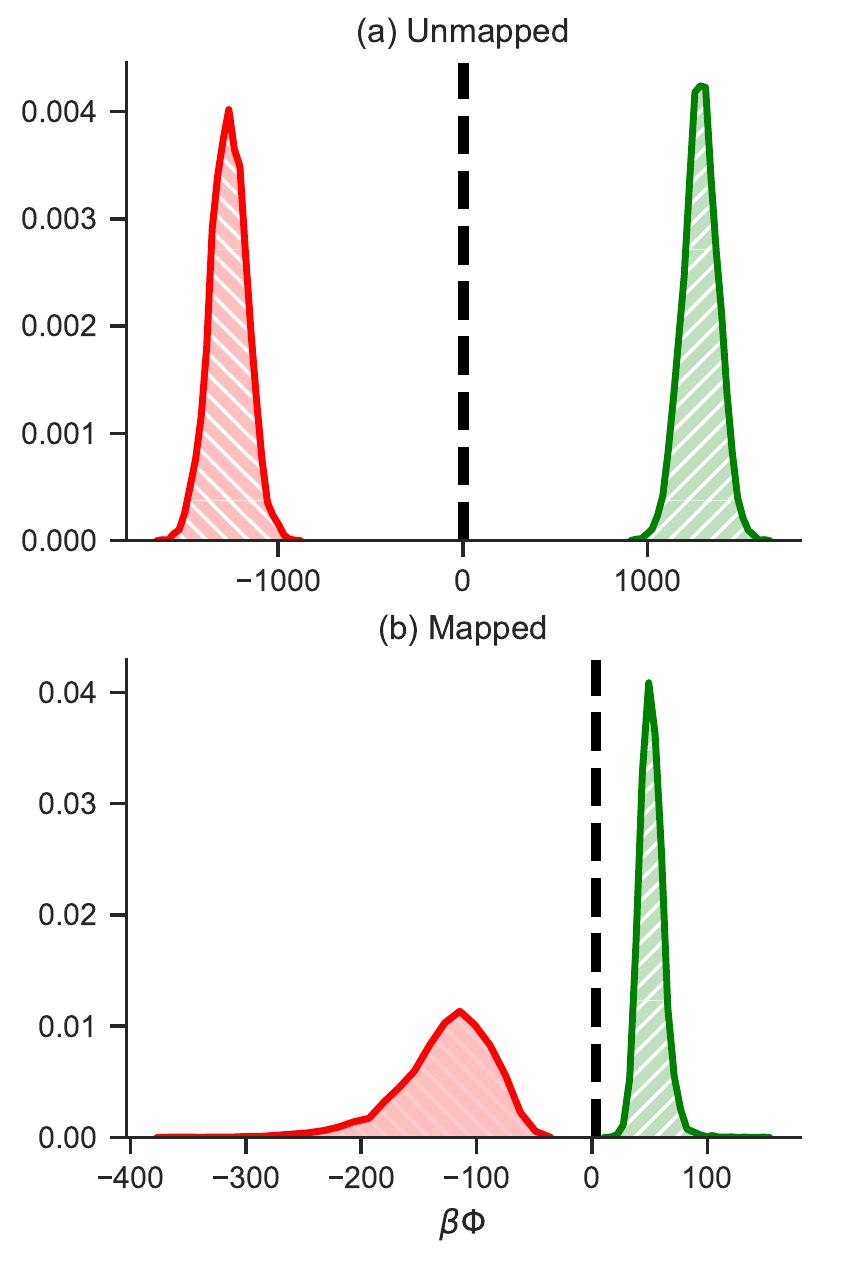}
\caption{
Normalized histograms of work (a) without and (b) with mapping between $\lambda_A = 14$ \AA ~and $\lambda_B = 18$ \AA. $\beta \Phi_F$ are colored green with forward diagonal hatches and $-\beta \Phi_R$ are colored red with backwards diagonal hatches.
\label{fig:phi4}}
\end{figure}

For this pair of states, TFEP fails to reproduce reference free energies because the work distributions are still not overlapping (Figure \ref{fig:phi4}). Without mapping, there is a gap of over 2000 $k_BT$ between the peaks of the forward and reverse work distributions. Mapping significantly reduces this gap, but there is not significant probability density at the free energy difference, making it difficult to estimate accurately.

\begin{figure}
\includegraphics[width=\figwidtho]{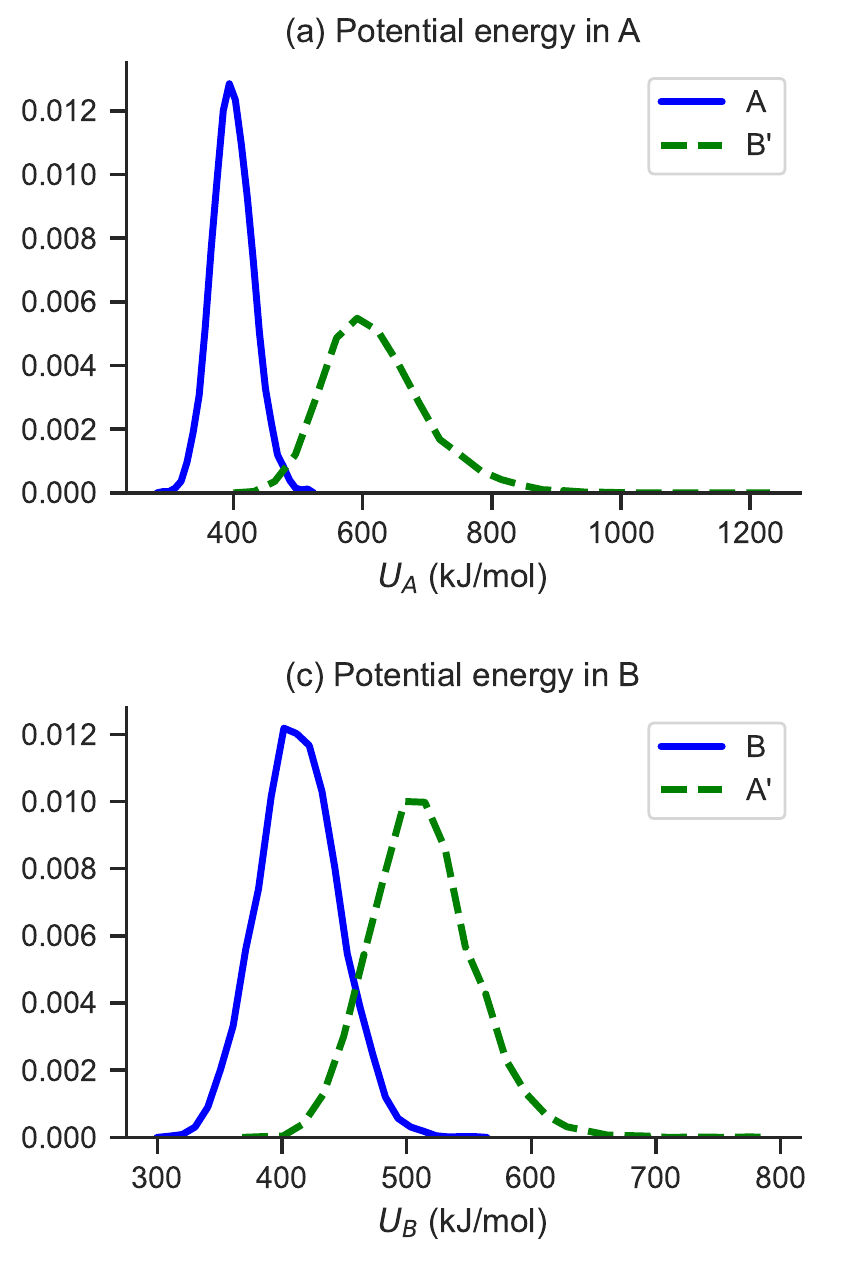}
\caption{
Normalized histograms of potential energies $U$ for (a) $\lambda_A = 14$ and (b) $\lambda_B = 18$ \AA, based on samples from the distribution (line) or mapped samples (dashed line).
\label{fig:U4}}
\end{figure}

The failure to achieve work distribution overlap occurs because mapping does not reproduce the targeted potential energy distribution (Figure \ref{fig:U4}). Overall, the mapped structures have a higher potential energy than the target distribution. For the forward mapping, the potential energy distribution is shifted higher and broadened. There is significant overlap with the target potential energy distribution. For the reverse mapping, there is a more pronounced shifting and broadening and the high-energy tail of the distribution is heavier. These suggest that the forward mapping is more successful than the reverse mapping.

\begin{figure}
\includegraphics[width=\figwidtho]{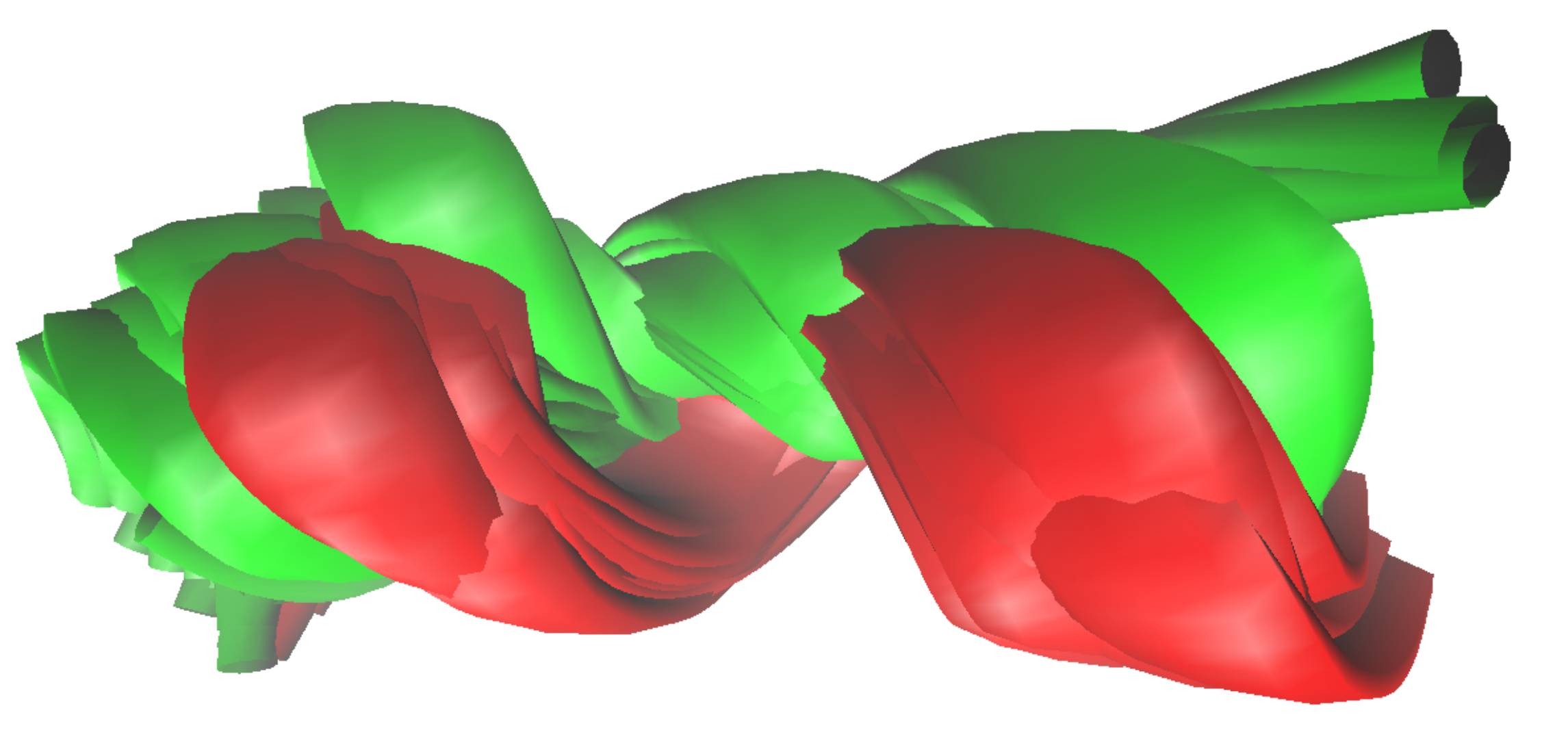}
\caption{
 Comparison of the targeted structures (red color) with the mapped structures (green color) in the reverse process from $\lambda_B = 18$ \AA~ to $\lambda_A = 14$ \AA.
\label{fig:mapped_da}}
\end{figure}

From a structural perspective, the failure of reverse mapping is related to the formation of hydrogen bonds. While state $A$ is characterized by a stable $\alpha$-helix, state $B$ is comparatively unfolded. Near the ends of the helix, structures mapped from state $B$ to state $A$ are similar to structures from state $A$ (Figure \ref{fig:mapped_da}). However, the middle of the helix is compressed and lacks hydrogen bonds. The mapped states are in a metastable state opposed to the most stable structure of the peptide.

Given the shortcomings of our more distant mappings, the development of improved training algorithms to  map between distant thermodynamic states across rugged potential energy surfaces is an intriguing area for future research. One possible approach to better training of mappings may be to borrow ideas from sampling. \citet{SbailoNoe21} demonstrated that learned mappings may be used to propose Monte Carlo moves between metastable states of a dimer. Mappings capable of generating reasonable structures in the target distribution will probably work well for TFEP.

As mentioned in the introduction, an alternative to mapping between distant molecular distributions is to map each molecular distribution to a comparable tractable distribution. Using this strategy, \citet{DingZhang20} were able to reproduce free energy differences between deca-alanine at very different temperatures, 300 K and 500 K. Subsequently, \citet{DingZhang21a} applied the same strategy to obtaining binding free energies for a host-guest complex. Reference calculations were performed with an attach-pull-release procedure with multiple intermediate states, suggesting that the end states occupy quite different configuration space. However, it is unclear whether their methods will work well on larger and more flexible protein-ligand systems. Moreover, as mentioned in the introduction, mapping between molecular distributions may be the only way to map subsets of coordinates.

\section{Conclusions\label{sec:conclusions}}

We have successfully used learned mappings in TFEP to reproduce reference free energy differences between conformations of a flexible bonded molecule with a rugged free energy surface, deca-alanine. The learned mappings were most successful when the unmapped states were similar. On the other hand, our procedure was unable to learn a mapping from an unfolded to alpha helical structure of the model peptide.

\section{Data Availability Statement}

Code to perform mapped free energy calculations and sample data are available at \url{https://github.com/swillow/jax_amber}.

\section{Acknowledgement}

We thank Yindong Chen and Joseph DePaolo-Boisvert for helpful discussions and comments on the manuscript. This research was financially supported in part by National Institutes of Health grant R01GM127712 to DDLM and National Science Foundation grants DMS 1916467 and DMS 2153029 to LK. The content is solely the responsibility of the authors and does not necessarily represent the official views of the National Institutes of Health or the National Science Foundation.

\bibliography{da_tfep}

\newpage

\setcounter{section}{0}

\renewcommand{\thesection}{\arabic{section}}   %%%% not here
%\appendix
\renewcommand{\thesection}{S\arabic{section}}    %%%% but here

%\section*{Supplementary Information}
\renewcommand{\thefigure}{S\arabic{figure}}
\setcounter{figure}{0}
\renewcommand{\thetable}{S\arabic{table}}
\setcounter{table}{0}

\renewcommand{\theequation}{S\arabic{equation}}
\setcounter{equation}{0}

\section*{Supplementary Material}

\begin{figure}
     \includegraphics[width=\figwidtho]{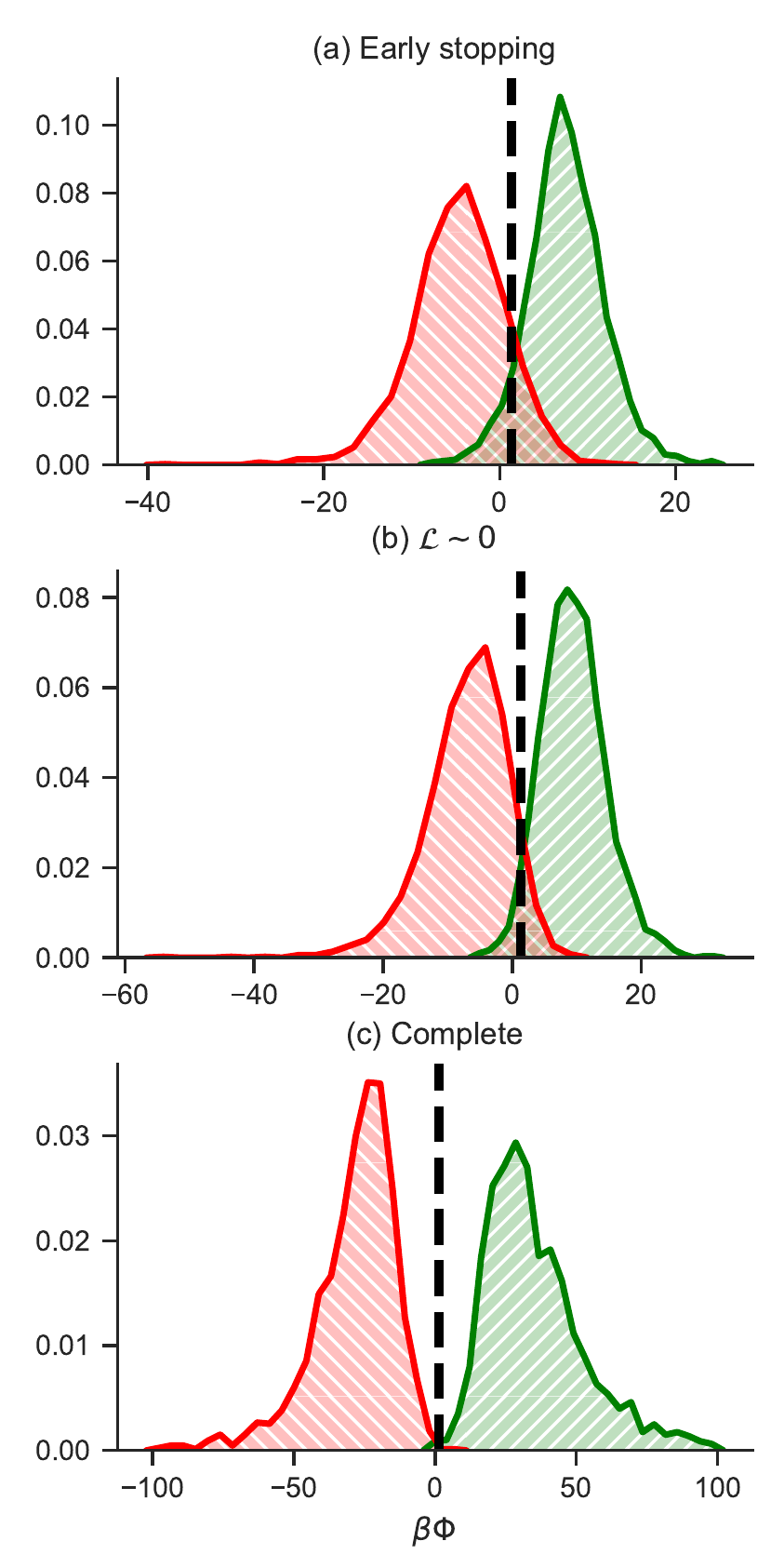} 
\caption{\label{fig:mapped_phi_test}Effect of stopping criterion on estimated probability densities of $\beta \Phi$ for test data between $\lambda_A = 20$ \AA~and $\lambda_B= 21$ \AA. Work distributions for mapping functions from (a) early stopping, (b) when the loss is closest to zero, and (c) after 150000 training steps. The reference free energy difference $\Delta F$ obtained via the MBAR estimator is represented by the vertical dashed line. A similar figure for the training data is available as Fig \ref{fig:mapped_phi}.}
\end{figure}

\begin{figure}
\includegraphics[width=\figwidtht]{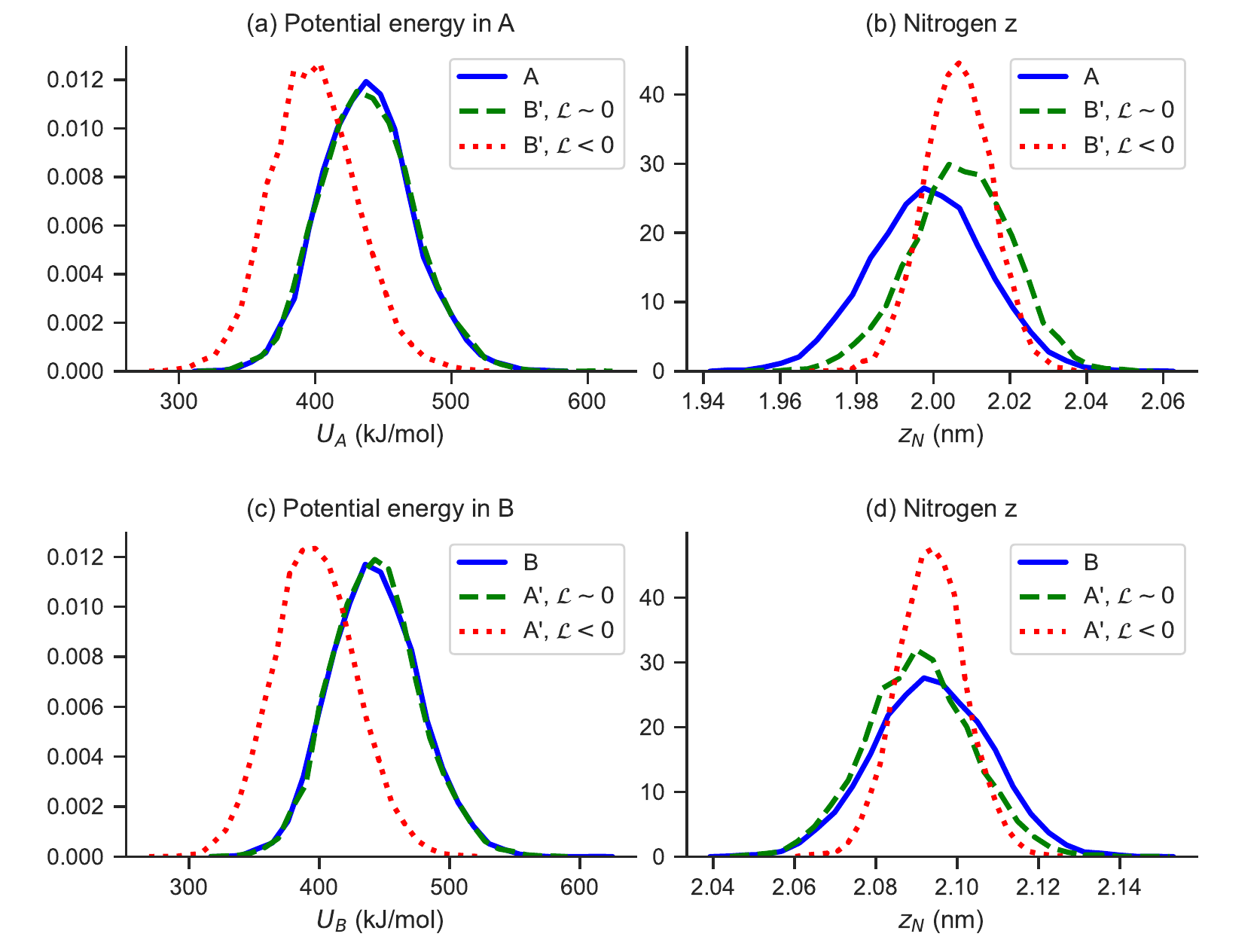}
\caption{
Estimated probability densities of (a) and (c) potential energies and (b) and (d) terminal nitrogen positions for $\lambda_A = 20$ \AA~and $\lambda_B = 21$ \AA. Normalized histograms are based on simulated (blue line) and mapped samples based on early stopping (green dashed line) and complete training (red dotted line). 
\label{fig:mapped_properties}}
\end{figure}

% We used $k=200$ kJ/mol/\AA$^2$ for the learned mapping, while $k=50$ kJ/mol/\AA$^2$ for the molecular dynamic simulations. Figure \ref{fig:Z} shows that when we minimized the training datasets with $k = 200$ kJ/mol/\AA$^2$, the mapped configurations were able to transform toward the targeted thermodynamic states. When we used $k=50$ kJ/mol/\AA$^2$ for the learned mapping, the mapped values $Z$ are far from the targeted thermodynamic states. 

\end{document}